\documentclass{PoS}
\usepackage{amssymb,amsmath,fontenc,times,mathptmx,ifpdf}
\usepackage{graphicx}
\usepackage{color}
\newcommand{\trh}{T_{\rm rh}}
\newcommand{\vrd}{\vr^{\dagger}}
\newcommand{\NCS}{N_{\rm CS}}
\newcommand{\nw}{N_{\rm w}}

\newcommand{\mueff}{\mu_{\rm eff}}
\newcommand{\beaa}{\begin{eqnarray*}}
\newcommand{\eeaa}{\end{eqnarray*}}
\newcommand{\bc}{\begin{center}}
\newcommand{\ec}{\end{center}}

\newcommand{\intvecx}{\int d^3 x\,}

\newcommand{\vecA}{{\bf A}}

\newcommand{\veck}{{\bf k}}

\newcommand{\vecp}{{\bf p}}

\newcommand{\vecx}{{\bf x}}       %(vette x)

\newcommand{\al}{\alpha}

\newcommand{\gm}{\gamma}
\newcommand{\dl}{\delta}

\newcommand{\lm}{\lambda}
\newcommand{\rh}{\rho}
\newcommand{\sg}{\sigma}
\newcommand{\ta}{\tau}

\newcommand{\ph}{\phi}
\newcommand{\vr}{\varphi}

\newcommand{\ps}{\psi}
\newcommand{\om}{\omega}

\newcommand{\Gm}{\Gamma}

\newcommand{\Sg}{\Sigma}
\newcommand{\Ph}{\Phi}

\newcommand{\half}{\frac{1}{2}}
\newcommand{\quart}{\frac{1}{4}}

\newcommand{\Tr}{\mbox{Tr}}
\newcommand{\tr}{\mbox{tr}}

\newcommand{\dmu}{\partial_{\mu}}

\newcommand{\psb}{\bar{\psi}}

\newcommand{\eela}[1]{\label{#1}\end{equation}}
\newcommand{\eeala}[1]{\label{#1}\end{eqnarray}}
\newcommand{\be}{\begin{equation}}
\newcommand{\ee}{\end{equation}}
\newcommand{\bea}{\begin{eqnarray}}
\newcommand{\eea}{\end{eqnarray}}

\newcommand{\mh}{m_{\rm H}}
\newcommand{\intveck}{\int \frac{d^3 k}{(2\pi)^3}\,}
\newcommand{\thalf}{\textstyle{\frac{1}{2}}}
\newcommand{\tquart}{\textstyle{\frac{1}{4}}}
\newcommand{\tsixth}{\textstyle{\frac{1}{6}}}

\PoS{PoS(LAT2005)022}

\title{Simulations in Early Universe Theory}

\ShortTitle{Simulations in Early Universe Theory}

\author{\speaker{Jan Smit}\\ %\thanks{Supported by FOM/NWO}\\
        Institute for Theoretical Physics, University of Amsterdam\\
        E-mail: \email{jsmit@science.uva.nl}}

\abstract{We give an impression of the type of results that
have been obtained with numerical lattice simulations of field
theory in the early universe.}

\FullConference{XXIIIrd International Symposium on Lattice Field Theory\\
         25-30 July 2005\\
         Trinity College, Dublin, Ireland}

\begin{document}

\section{Introduction}
\begin{figure}
\includegraphics[width=\textwidth]{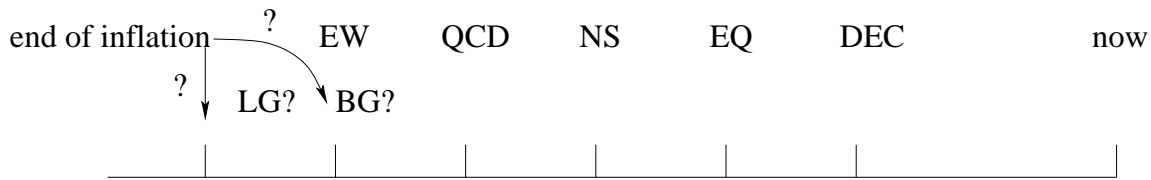}
\caption{Transitions in the early universe. Going backwards in time
there was the decoupling of photons (DEC), observed today as the
Cosmological Microwave Background, a point of equal energy in
`radiation' and (non-relativistic) `matter' (EQ), a period of
nucleosynthesis (NS), probably a QCD transition, and presumably an
ElectroWeak transition. There are good arguments for a period of
inflation, with support from the microwave background. When
inflation ended is uncertain. Leptogenesis (LG) and/or
baryogenesis (BG) is supposed to have taken place after inflation
and before or during the EW transition.}
\label{timeline}
\end{figure}

When indicating various transitions that are supposed to have
taken place in the early universe
\cite{KT}
on a time line (Fig.\ \ref{timeline}),
%
%LG: leptogenesis \hfill EW \& QCD transitions\\
%BG: baryogenesis\\
%NS: nucleosynthesis\\
%EQ: equal energy in `radiation' and `matter'
%\hfill
%DEC:  photon decoupling ($\leftrightarrow$ CMB)
%
one is confronted by the fact that what we know with confidence
goes back only to the time of nucleosynthesis. Was there a QCD
transition? Probably, because its temperature of 150 MeV is only
two orders of magnitude higher than that at the start of
nucleosynthesis. The temperature of the conventional electroweak
transition is already three orders of magnitude higher than that
of the QCD transition. Could it have been different? Where should
one put the end of inflation, baryogenesis? To make progress we
need to understand as much as possible of the dynamical behavior
of a universe, given our knowledge of fundamental interactions.
This often concerns non-perturbative phenomena.

This talk is about numerical simulations in field theory, for
early times, around and before the supposed electroweak
transition, using lattice regularization. We will give an
impression by examples of what has been done, it is not a review.

\section{Inflation, preheating, defects, baryogenesis,
thermalization}
Let us take a brief look at various topics in the
title of this section.
\subsection*{Inflation}
\begin{figure}
\includegraphics[width=0.5\textwidth]{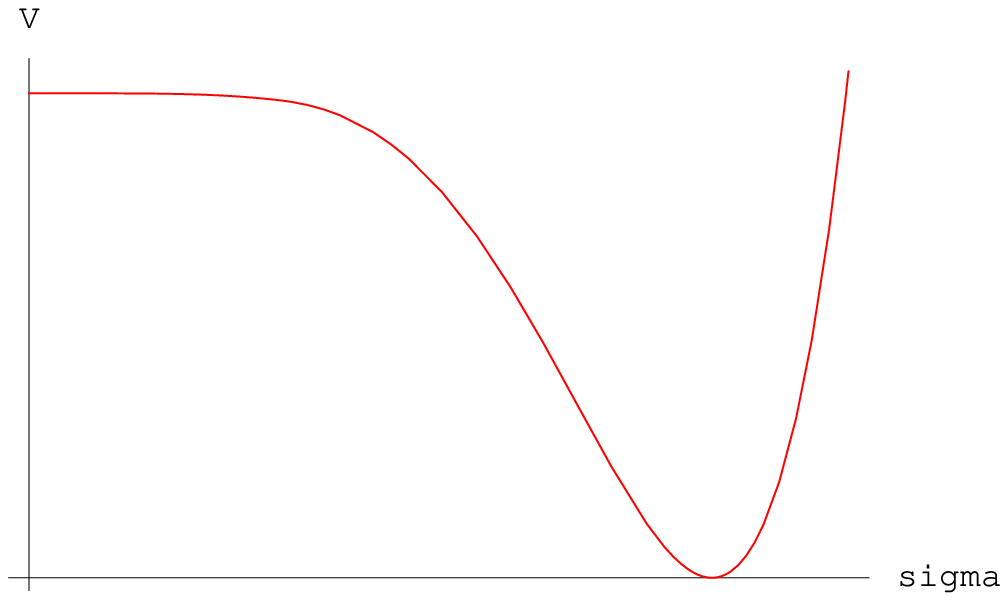}
\includegraphics[width=0.5\textwidth]{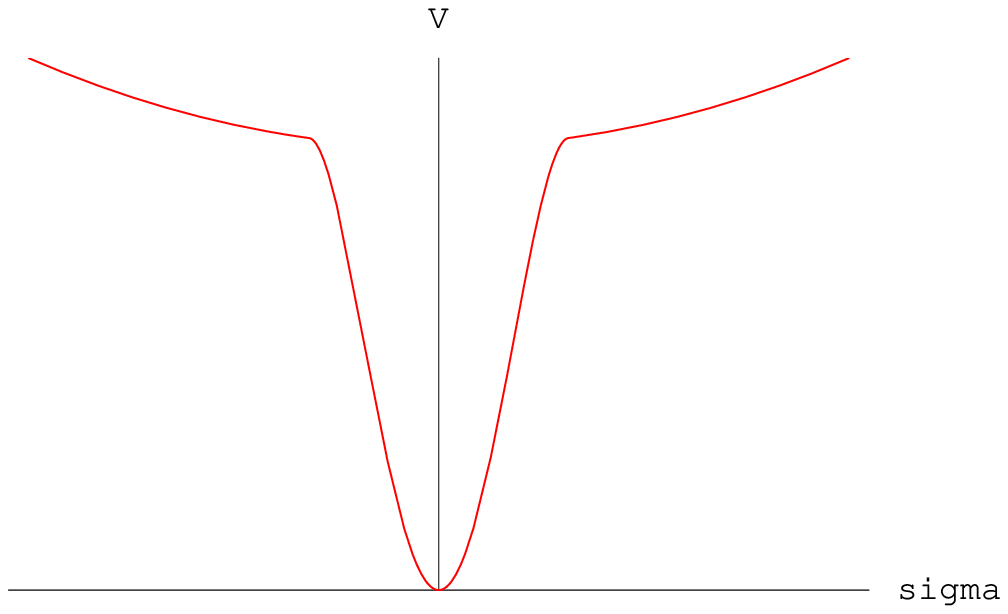}
\caption{Small-field (left) and large-field (right) inflaton potentials.}
\label{inflatonpots}
\end{figure}

Simple models \cite{LL}
are formulated in terms of an inflaton
field,
$\sg$, the Friedmann-Lema\^itre-Robertson-Walker scale factor $a$,
which satisfy the dynamical equations
\beaa
0&=& \ddot\sg + 3H\dot\sg-\frac{\nabla^2\sg}{a^2}+
\frac{\partial V(\sg,\cdots)}{\partial\sg},
\\
H \equiv\frac{\dot a}{a}&=& \sqrt{\frac{\rh}{3m_{\rm P}^2}},
\\
\rh &=& \frac{\dot\sg^2}{2} + \frac{(\nabla\sg)^2}{2a^2} +
V(\sg,\cdots) + \cdots
\eeaa
in flat space. Here $H$ is the Hubble rate and $\rh$ the energy
density. When the potential energy of the inflaton dominates and
its potential is very flat, there is strong damping and `slow
roll', a nearly constant Hubble rate and approximately exponential
expansion
$a\propto e^{Ht}$.
%$a\propto e^{\int^t H}$.
Inflaton models can be divided into two classes: small-field and
large-field models (Fig.\ \ref{inflatonpots}), depending on $\sg$
rolling away or towards the origin, respectively. Other names are
`new inflation' and `chaotic inflation', respectively.
The large-field potential shown in Fig.\ \ref{inflatonpots} (right)
is an example of a so-called hybrid-inflation model:
\be
V(\sg,\vr) = V_0+ \half\,\mu_\sg^2\sg^2 + g^2\sg^2\vrd\vr
-\mu^2\vrd\vr + \lm(\vrd\vr)^2,
\label{hybridpot}
\ee
where $\vr$ could be a Higgs field. Flatness of the potential
requires relatively small $\mu_\sg$, typically $\mu_\sg \ll \mu$.
During inflation the inflaton field is large,
$g^2\sg^2 - \mu^2 >0$ and $\vr=0$ (read $\langle\vr\rangle = 0$).
Inflation ends when $g^2\sg^2 - \mu^2$ goes negative. Assuming the $\vr$
dynamics to be much faster than that of the inflaton, we can solve for
$\vrd\vr$ and substitute it back into the potential:
$\vrd\vr \to (\mu^2 - g^2\sg^2)/\lm$. This gives the
trough near the origin shown in the right plot of Fig.\
\ref{inflatonpots} (a similar substitution has been assumed in the
left plot). When the inflaton has gone down the `waterfall', the
universe is assumed to be cold\footnote{This is different in `warm
inflation', see e.g\ \cite{Bastero-Gil:2004tg}.} and without
appreciable baryon- and lepton-number densities, because it has
just undergone a huge expansion. The energy in the inflaton has to
be transferred to other degrees of freedom (d.o.f.) in order for
baryogenesis to become possible, such that it ends up eventually
at the standard hot radiation-dominated stage during
nucleosynthesis. This heating up (called `re-heating') might take
too long in a given model, but non-perturbative pre-heating
processes may have helped to achieve it.

\subsection*{Preheating}
Rewriting the potential in the form
$V(\sg,\cdots) = V(\sg) + (g^2 \sg^2-\mu^2)\vrd\vr + \cdots$ reveals an
effective $\vr$ mass parameter,
\[
\mueff^2 =-\mu^2 + g^2 \sg^2.
\]
Its time dependence can have drastic non-perturbative effects:

$\cdot$ $\langle\sg\rangle$ oscillates $\to$ %parametric resonance
$\vr$ resonates \cite{Kofman:1997yn};

$\cdot$ $\mueff^2 <0$ $\to$ tachyonic instability \cite{Felder:2000hj}.

Also the $\sg$ modes can resonate and $\sg$ is also tachyonic during the time
that $\partial^2 V/\partial\sg^2 < 0$. These non-perturbative effects
can result in large occupation numbers of $\sg$ and $\vr$, in a limited
momentum range. Such processes are called pre-heating.

\subsection*{Defects}
Preheating processes often result in the formation of defects,
depending on the fields involved:
\begin{itemize}
\item[]
domain walls, strings, monopoles (see e.g.\ \cite{KT}), textures
\cite{Turok:1989ai}, Q-balls, I-balls, oscillons (see e.g.\
\cite{Broadhead:2005hn,Farhi:2005rz}), half-knots, Chern-Simons
number-densities,
\ldots
\end{itemize}
They may play a role in baryogenesis, thermalization and structure formation.

\subsection*{Baryogenesis}

The aim is to explain the baryon to photon ratio
$n_B/n_\gm \simeq 6.5\, 10^{-10}$
from $n_B=0$ (inflation) and B-number, C \& CP violation, during
particle physics processes out of equilibrium. Many scenarios for
doing this have been proposed, see e.g.\
\cite{Dine:2003ax,Buchmuller:2005eh}. Here we shall focus on one
for which lattice simulations have been carried out:
%\subsection*{Cold ElectroWeak Baryogenesis}
Cold ElectroWeak Baryogenesis
\cite{Garcia-Bellido:1999sv,Krauss:1999ng,Copeland:2001qw,Garcia-Bellido:2003wd,Tranberg:2003gi}.

In this scenario one assumes that the electroweak transition was tachyonic
of the type discussed below \eqref{hybridpot}, and occurred
at the end of a period of inflation when the universe was still cold
(so the electroweak transition was {\em not}
caused by the falling temperature of the universe.)
%So $\vr$ is the usual Higgs field of the
%Standard Model, which has been minimally extended with a gauge-singlet
%inflaton $\sg$.
The electro-weak anomaly in the baryon current
\[
\dmu j^{\mu}_B = 3\,
q, \quad
q= \frac{1}{16\pi^2}\, \tr F_{\mu\nu}\tilde F^{\mu\nu}
%= \dmu j^{\mu}_{\rm CS}
\]
with $q$ the `topological-charge' density of the $SU(2)$ gauge field,
can be integrated over the transition to give a baryon asymmetry
\be
B(t) = 3\int_0^t dt'\, \intvecx
\langle q(\vecx,t')\rangle
%\langle \dnot j^0_{\rm CS}(\vecx,t')\rangle
%$\\$
%= 3\langle\ncs(t) - \ncs(0)\rangle
.
\label{anomaly1}
\ee
This can be non-zero because the tachyonic transition occurred under the
influence of C and CP violation.

\subsection*{Thermalization}
The important question is: how long does it take to thermalize the
expanding universe after inflation? The `re-heating' temperature
$T_{\rm rh}$ marks the beginning of the radiation-dominated
universe. Nucleosynthesis constrains this to $T_{\rm rh}\gtrsim 4$
MeV \cite{Hannestad:2004px}. Thermalization is a non-perturbative
problem in out-of-equilibrium quantum field theory.

\section{Classical approximation}
Non-perturbative computations of expectation values observables like
\[
\langle O(t)\rangle = \Tr\, \rh\, O(t) =
\Tr\, \rh\, U^\dagger(t)\, O\, U(t)
\]
where $\rh$ is the density operator and $U(t)$ is the evolution
operator in {\em real time}, are very difficult in quantum field
theory. The classical approximation can therefore be very useful,
since it is possible to solve the fully non-linear field equations
on a computer (a perturbative analysis is given in
\cite{AaSm97,Aarts:1998kp}). To formulate this we can think of a coherent state
\cite{Salle:2000hd} or Wigner representation
\cite{Mrowczynski:1994nf}
\[
\langle O(\ph,\pi)\rangle = \int D\ph_{\rm c}\, D\pi_{\rm c}\,
\rh_{\rm c}(\ph_{\rm c},\pi_{\rm c})\, O(\ph_{\rm c},\pi_{\rm c})
\]
where $\ph_{\rm c}$ and $\pi_{\rm c}$ are classical conjugate variables.
If, under suitable
conditions, the functional $\rh_{\rm c}$ is positive
(generally not true in the quantum regime), then we can
use it as a probability
for initial conditions. The classical approximation then goes as follows:

%\begin{enumerate}
%\item
\indent
1. draw an initial configuration from $\rh_{\rm c}(\ph_{\rm c},\pi_{\rm c})$\\
%\item
\indent
2. solve classical e.o.m.\ for $\ph_{\rm c}$ and $\pi_{\rm c}$\\
%\item
\indent
3. evaluate $O(\ph_{\rm c},\pi_{\rm c})$\\
%\item
\indent
4. average over initial configurations
%\end{enumerate}

\noindent
One expects a classical approximation to be applicable in cases where
large occupation numbers dominate, which depends on the observable and on the
initial conditions. Let us take a look at occupation numbers. For a free
scalar field the usual definition is
\[
n_\veck = \langle a_\veck^\dagger a_\veck\rangle,
\quad
a_\veck = \frac{1}{\sqrt{2\om_\veck}}(\om_\veck \ph_\veck + i \pi_\veck),
\]
where $\ph_\veck$ and $\pi_\veck$ are Fourier modes of the
canonical variables, and $\om_k$ is the corresponding energy
($\sqrt{k^2 + m^2}$). We can generalize this to time-dependent
quasi-particle energies $\om_\veck$ and occupation numbers
$n_\veck$, $\tilde n_\veck$, defined by
%\beaa
%\langle \ph_\veck\ph_\veck^\dagger\rangle
%&\equiv& \left(n_\veck + \half\right)\frac{1}{\om_\veck},
%\quad
%\langle \pi_\veck\pi_\veck^\dagger\rangle
%\equiv \left(n_\veck + \half\right)\om_\veck
%\\
%\langle \ph_\veck\pi_\veck^\dagger\rangle
%&\equiv&
%\tilde n_\veck + \frac{i}{2}.
%\eeaa
\be
\langle \ph_\veck\ph_\veck^\dagger\rangle
\equiv \left(n_\veck + \half\right)\frac{1}{\om_\veck},
\quad
\langle \pi_\veck\pi_\veck^\dagger\rangle
\equiv \left(n_\veck + \half\right)\om_\veck,
\quad
\langle \ph_\veck\pi_\veck^\dagger\rangle
\equiv
\tilde n_\veck + \frac{i}{2}.
\label{partnumbers}
\ee
Given the correlation functions, we can easily solve for
$\om_\veck$, $n_\veck$ and $\tilde n_\veck$,
and see if $n_\veck$ and $\tilde n_\veck$ are larger than 1.

Consider now a tachyonic transition at $t=t_c$, when
$\mueff^2(t_c)=0$. Assume that the couplings are weak enough that
a gaussian approximation makes sense. The e.o.m.\ for a
component of $\vr$ is then
\[
%\ddot \ph -\nabla^2\ph + \mueff^2\, \ph = 0
\ddot \ph_\veck + (\mueff^2+ k^2) \ph_\veck = 0.
\]
Two cases for which this equation can be solved are
\cite{Asaka:2001ez,Garcia-Bellido:2002aj}
\be
\mueff^2 = - M^3 (t-t_c),
\quad  M^3 \equiv -2g^2\, \sg_c\dot\sg_c,
\label{nonquench}
\ee
valid near $t_c$, and the sudden quench \cite{Smit:2002yg}
\bea
\mueff^2 &=& +\mu^2, \quad t<t_c,
\\
&=&  - \mu^2, \quad t>t_c.
\label{quench}
\eea
One finds that
$n_k$ and $\tilde n_k$
grow faster than exponential\footnote{For the quench,
$n_k\sim\exp[\sqrt{\mu^2-k^2}(t-t_c)]$.}
and $n_k+1/2-\tilde n_k \to 0$,
for $k < k_{\rm max}$:
\bea
k_{\rm max} &=& \sqrt{M^3 (t-t_c)}
\label{kmaxGB}
\\
&=& \mu, \quad \mbox{for the quench.}
\label{kmaxST}
\eea
Quantum $\langle\cdots\rangle$
dominated by the growing modes
can after some time be well reproduced by the classical distribution
\be
%\exp\left[-\half\sum_{\veck} \mbox{}^{\prime}
\exp\left[-\half\sum_{k \leq k_{\rm max}}
\left(\frac{|\xi^+_\veck|^2}{n_k + 1/2 + \tilde n_k}
+ \frac{|\xi^-_\veck|^2}{n_k + 1/2 - \tilde n_k}\right)\right],
\quad
\xi^{\pm}_{\veck} =
\frac{1}{\sqrt{2\om_k}}\left(\om_k\,\ph_{\veck} \pm \pi_\veck\right).
\label{gaussensemble}
\ee
%where
%\[
%\xi^{\pm}_{\veck} =
%\frac{1}{\sqrt{2\om_k}}\left(\om_k\,\ph_{\veck} \pm \pi_\veck\right),
%\]
%and
%$\sum\mbox{}^{\prime}$: $k \lesssim k_{\rm max}$.
The condition $k \lesssim k_{\rm max}$ can be relaxed provided
that the high-momentum
modes can be dealt with by renormalization.

So for tachyonic transitions we obtain well motivated initial conditions for a
classical approximation:

\indent
1a. at time $t_{\rm i} > t_c$ when $n_k,\tilde n_k \gg 1$ and non-linear terms
in the e.o.m.\ are still small\\
\indent
1b. (`just the half') alternatively, choose $t_{\rm i} = t_c$\\
\indent
2. choose between
$k_{\rm max}$ as in (\ref{kmaxGB}) or (\ref{kmaxST}),
%(hence $k_{\rm max} \ll \pi/a$, $a =$ lattice spacing),
or $k_{\rm max} = $ `cutoff' (all modes)\\
\indent
3. draw the initial $\ph$ and $\pi$ from the gaussian ensemble
\eqref{gaussensemble}

\noindent
The alternative 1b.\ is possible because quantum and classical
evolution are formally identical for a free field. The initial
particle numbers are zero in this case, so in
\eqref{gaussensemble}, $n_k + 1/2 \pm \tilde n_k = 1/2$ whence the
name `just the half' \cite{Smit:2002yg}. In case of $k_{\rm
max}\neq$ `cutoff', the initial modes with $k> k_{\rm max}$ are
set to zero. Proposals of this kind appeared earlier in the
literature
\cite{Mrowczynski:1994nf} (see also
\cite{Yavin}).

%\subsection*{Examples}
An attractive formulation for numerical simulations is obtained by
first discretizing the action on a space-time lattice, with
spacings $a_t\ll a_s$, and then deriving the field equations from
the stationary action principle. This especially useful for gauge
fields \cite{Ambjorn:1991pu}, and it leads to leapfrog-type e.o.m.
Some milestones in early universe simulations are the computation
of the sphaleron rate at high temperature
\cite{Ambjorn:1995xm,Bodeker:1999gx}, and the unveiling of
`re-scattering' effects after parametric resonance
\cite{Khlebnikov:1996mc}
and tachyonic preheating \cite{Felder:2000hj}. In the following
some examples will be mentioned of more recent work.

\begin{figure}
\includegraphics[width = 0.5\textwidth]{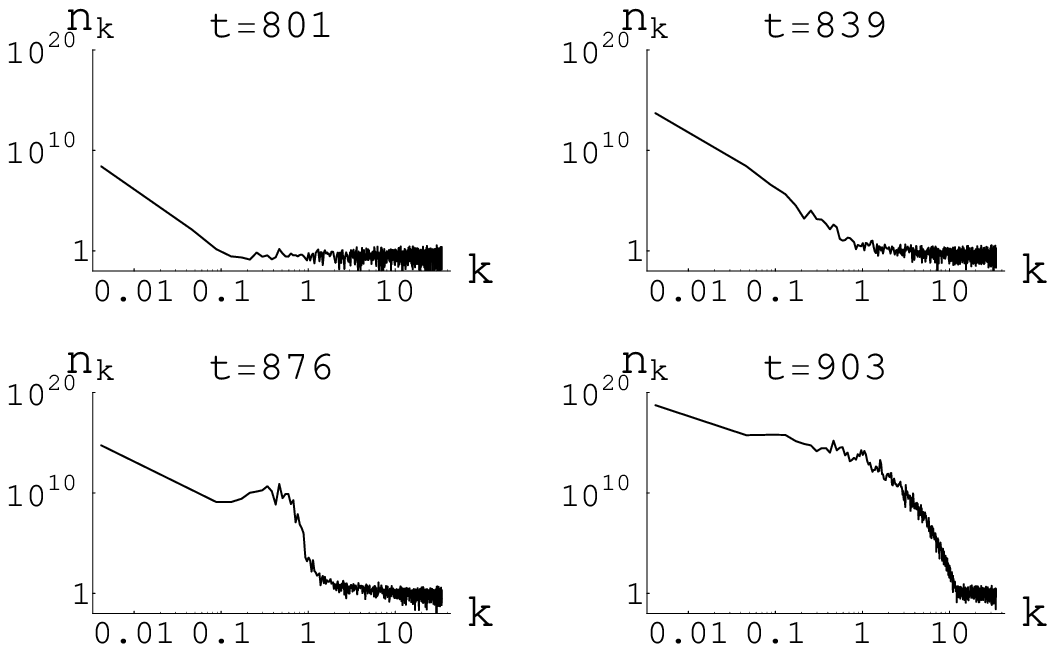}
\includegraphics[width = 0.5\textwidth]{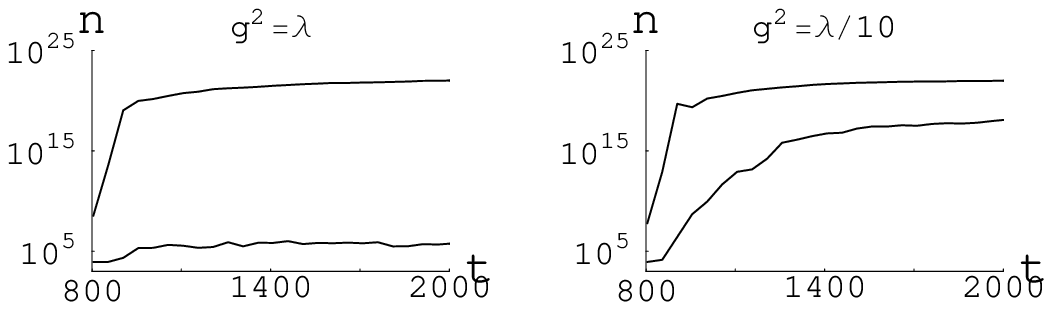}
\caption{Inflaton occupation numbers $n_k$ at various times. A resonance
is visible at $t=876$ ($t$ and $k$ in $\sqrt{\lm} v$-units). From
\cite{Desroche:2005yt}.}
\label{desroche1}
\end{figure}

\subsubsection*{Example: particle numbers in `New inflation'
\cite{Desroche:2005yt}}
The inflaton potential is a small field model of the Coleman-Weinberg form
\[
V(\sg) =  \quart\,\lm\sg^4\left(\ln\frac{|\sg|}{v} - \quart\right)
+ \frac{1}{16}\, \lm v^4.
\]
Using an equilibrium one-loop potential for non-equilibrium dynamics is
questionable, in principle, but it may be justifiable by assuming relatively
fast dynamics for the non-inflaton d.o.f. The parameters are
$\lm = 10^{-12}$, $v=10^{-3}\, m_{\rm P}$ ($m_{\rm P}$ is the Planck mass),
typical values for satisfying the
CMB constraints. The simulation takes the Hubble expansion into account.
Initially $\langle \sg\rangle\approx 0$, after inflation
it rolls into minimum of $V$ where it oscillates, the energy decays into
$\sg$-particles. Figure \ref{desroche1} (left)
shows how inflaton particle-numbers have become very large due to
tachyonic preheating, and one can see also parametric resonance
around time 876.

By including another field $\ph$ the transfer of energy to other d.o.f.\
is studied. The potential is taken as
$V=V(\sg) + \half\, g^2 \sg^2\ph^2$. This is non-tachyonic for
$\ph$ -- the $\ph$ particle numbers grow exponentially because of parametric
resonance.
Figure \ref{desroche1} (right) shows the number densities
$n=\intveck n_k$ for the two species, for two values of the couplings.
The resonance is more efficient for smaller $g^2$, but
$n_\ph \ll n_\sg$ and the inflaton does not loose its energy very fast.
So in this case preheating is not very efficient and the reheating
temperature is estimated analytically to be
$\trh \approx \sqrt{m_{\rm P}\,\Gm_{\mbox{}\!\sg\to\ph\ph}}$
($\approx 10^7$ GeV for $g^2 \lesssim \lm$).
% where $\Gm_{\mbox{}\sg\to\ph\ph}$ is the inflaton decay rate.

\subsubsection*{Example: mixing in an inflaton--SU(2)-Higgs model
\cite{Garcia-Bellido:2002aj}}
\begin{figure}
\includegraphics[width = 0.5\textwidth]{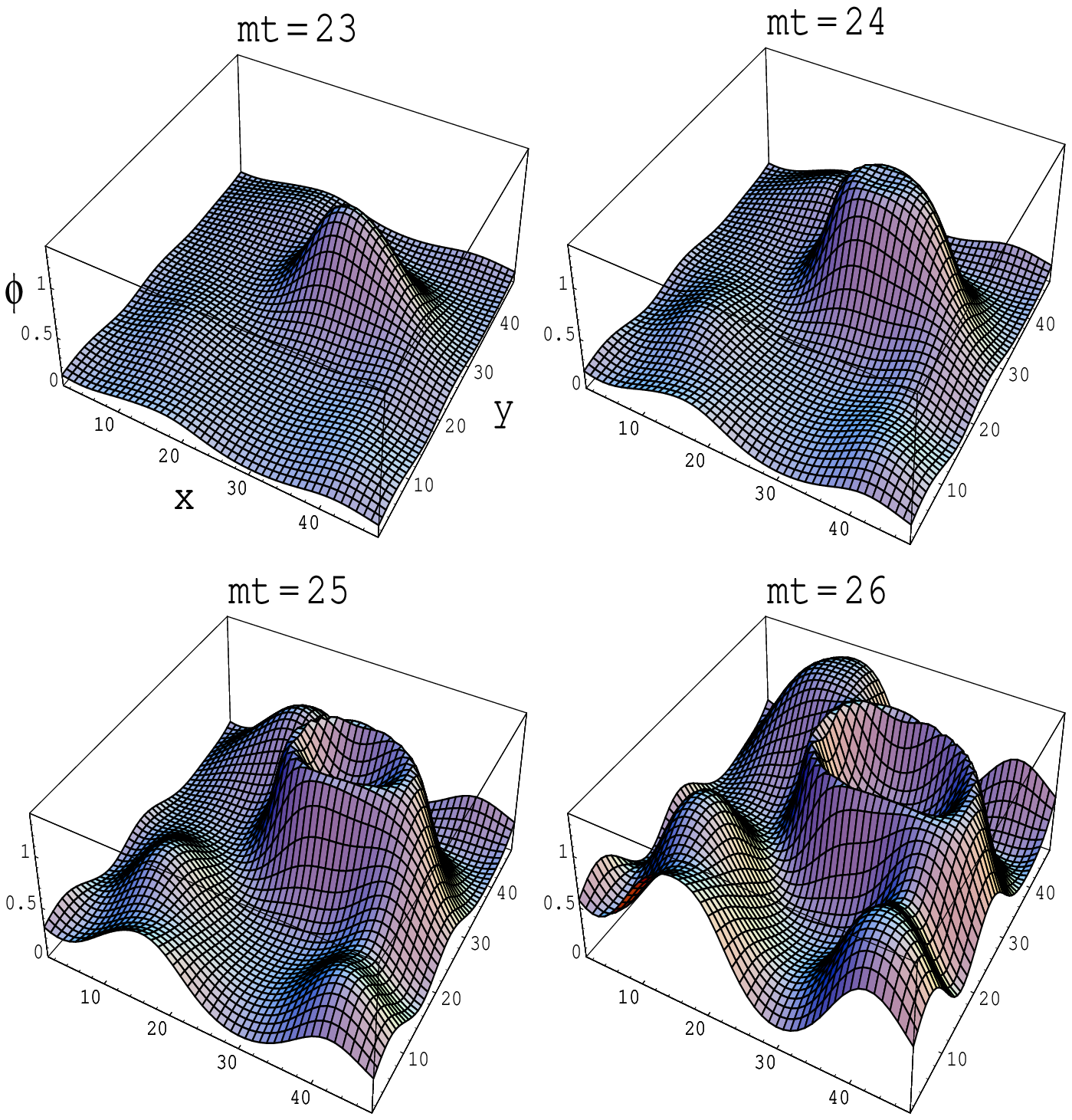}
\includegraphics[width = 0.5\textwidth]{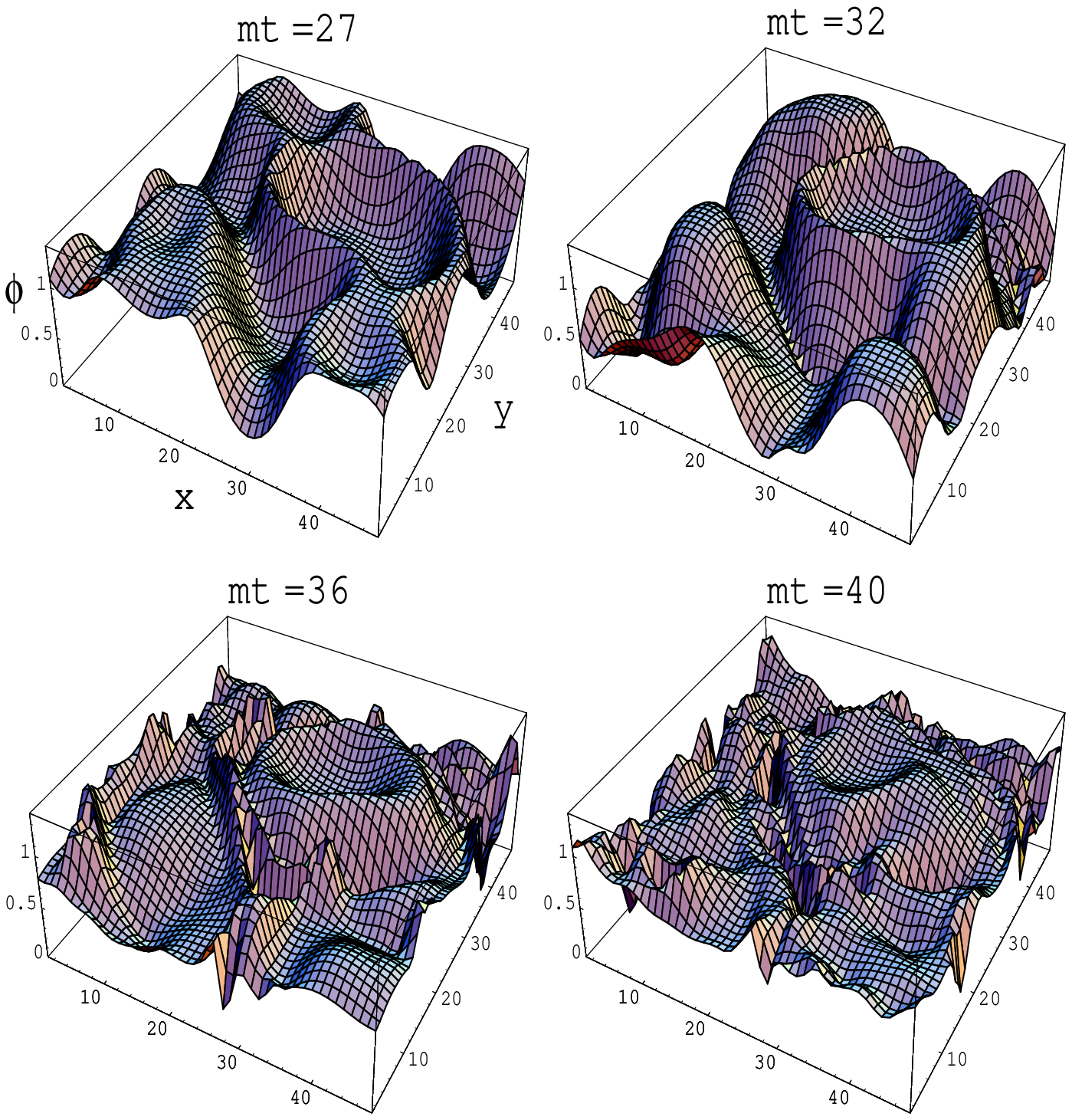}
\caption{Higgs-field amplitude in $x$-$y$ slice at various times.
From \cite{Garcia-Bellido:2002aj}.}
\label{GB1}
\end{figure}

The potential corresponds to a large-field model
%\[
%V(\sg,\vr) = \half\,\mu_\sg^2 \sg^2 + g^2\sg^2\vrd\vr +
%\lm(\vrd\vr - \half\,v^2)^2,
%\]
\be
V=\textstyle{\half}\,\mu_\sg^2\sg^2+g^2\sg^2 \vrd\vr +
\lm\left(\vrd\vr-\textstyle{\half}\,v^2\right)^2,
\label{inflHiggs}
%\vr = r\, e^{i\al}/\sqrt{2}
\ee
in which the inflaton is coupled to the Higgs sector of the
Standard Model, albeit with somewhat unrealistic couplings,
$\lm=0.11/4$, $g^2 =2\lm$. The Hubble expansion is negligible for
$v=246$ GeV ($H=O(10^{-14})$ GeV).
The initial conditions where drawn from a fit to the
$k$-dependence of the distribution at $t=t_{\rm i}$, with $M=
0.18\, m$, $m=\sqrt{\lm}\, v$ (cf.\ (\ref{nonquench})). Figure
\ref{GB1} shows the Higgs field in an $x$-$y$ slice through the
lattice at various times after the transition. Intuitively, the
non-linear scattering of waves causes a good mixing of field modes
and shortens the period needed to reach the thermalized state.

\subsubsection*{Example: equation of state in an inflaton--U(1)-scalar
model \cite{Borsanyi:2003ib}} The equation of state (e.o.s.),
i.e.\
$p=p(\rh)$, is a basic ingredient in hydrodynamic descriptions of
the early universe. In simple descriptions the cosmic fluid is
often approximated by mixture of various contributions, such as
`radiation' $p=\rh/3$, `matter' $p\simeq 0$, `vacuum' $p=-\rh$,
and it is interesting to see if and when such an approximation
makes sense. The potential in this study is the U(1) version of
\eqref{inflHiggs}. The energy scale is presumed much higher in
this case, with parameters
$g^2=10^{-2}$, $\lm = g^2/2$, $\mu_\sg=4\,10^{11}$ GeV,
$m_{\rm H}=\sqrt{2}\, m = 9\, 10^{14}$ GeV. There are no gauge
fields. The Hubble expansion is taken into account. Writing
\[
\vr = r\, e^{i\al}
\]
the pressures of the inflaton, Higgs and Goldstone bosons are
defined as
\beaa
p_{\rm H} &=&
%\frac{\dot r^2}{2} - \frac{(\nabla r)^2}{6 a^2} +
%\frac{\lm(r^2-v^2)^2}{4}
%,\quad \vr = r\, e^{i\al}
%\half \,\dot r^2 - \frac{1}{6}\,
\thalf\,\dot r^2 - \tsixth\,
%\left(\nabla r/a\right)^2 +
(\nabla r/a)^2 +
\tquart\,\lm(r^2-v^2)^2,
\\
p_{\rm G} &=&
%\frac{r^2\dot\al^2}{2} - \frac{(\nabla \al)^2}{6a^2}
%\half\, r^2\dot\al^2 - \frac{1}{6}(\nabla \al/a)^2
\thalf\, r^2\dot\al^2 - \tsixth(\nabla \al/a)^2,
\\
 p_\sg &=&
 %\frac{\dot\sg^2}{2} - \frac{(\nabla\sg)^2}{6a^2} +
%\frac{(m_\sg^2+g^2r^2)\sg^2}{2}
%\half\,\dot\sg^2 - \frac{1}{6}(\nabla\sg/a)^2 +
\thalf\,\dot\sg^2 - \tsixth(\nabla\sg/a)^2 +
\thalf\,(\mu_\sg^2+g^2r^2)\sg^2.
\eeaa

\begin{figure}
\bc
\includegraphics[width = 0.7\textwidth]{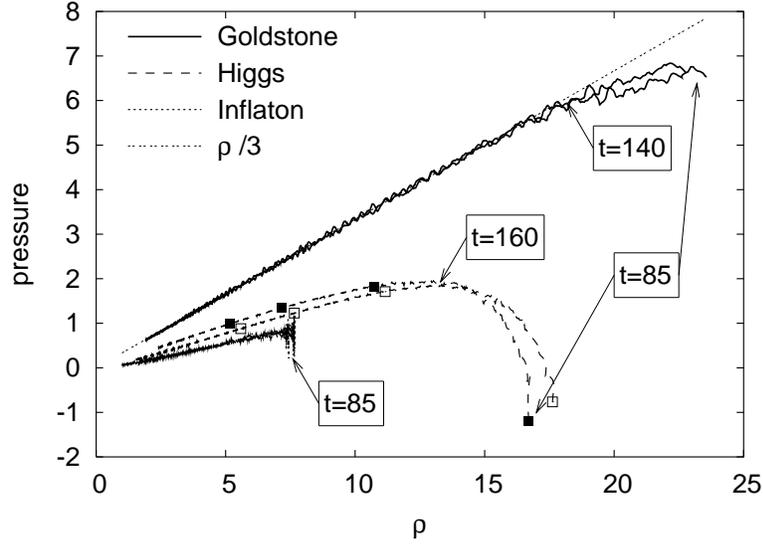}
\ec
\caption{Parametric plot of the pressures $p_{\rm H}$, $p_{\rm G}$
and $p_\sg$ vs the energy density $\rh$, for times $t m=85 - 200$,
and two lattice spacings, $a_s m = 0.5$, 1. From
\cite{Borsanyi:2003ib}.}
\label{Borsanyi1}
\end{figure}

Figure
\ref{Borsanyi1} shows the emergence of an equation of state and
its break-up into various contributions. We see how the Goldstone
modes approach the radiation e.o.s.\
$p=\rh/3$, whereas the massive Higgs and inflaton modes clearly
differ from this. The complete pressure is the sum of these
components and this study indicates that a definite e.o.s.\
emerges around $tm \approx 85$ after the transition. Some of the
contributions were interpreted in terms of global strings
\cite{Borsanyi:2003ib}.

\subsubsection*{Example: strings and hot spots
\cite{Copeland:2002ku}}
Global strings are an example of (unstable) topological defects
which one expects to be produced in a tachyonic transition, and
which may be relevant for elucidating the mechanism of tachyonic
pre-heating. Other possible objects of interest are `hot spots',
places where the inflaton field has `jumped' over the trough in
Fig.\ \ref{inflatonpots} (right) and ends up again at values
$|\sg| > \sg_c$. The model is the same U(1) inflaton-`Higgs' model
as in the previous example (with $v^2/2\to v^2$).
The parameters are
$g^2 = 10^{-4}$, $\lm = 10^{-2}$,
$m\equiv \sqrt{2\lm}\, v = 3\, 10^{15}$ GeV, and the expansion
rate is taken to have the constant value $H=3.6\,10^{-3}\, m$. The
initialization is at $\sg=\sg_c$ when
$\mueff^2 \equiv g^2 \sg^2 -m^2 = 0$, \`a la `just the half' with
$k_{\rm max}$ = `cutoff'.

\begin{figure}
\includegraphics[width = 0.5\textwidth,clip]{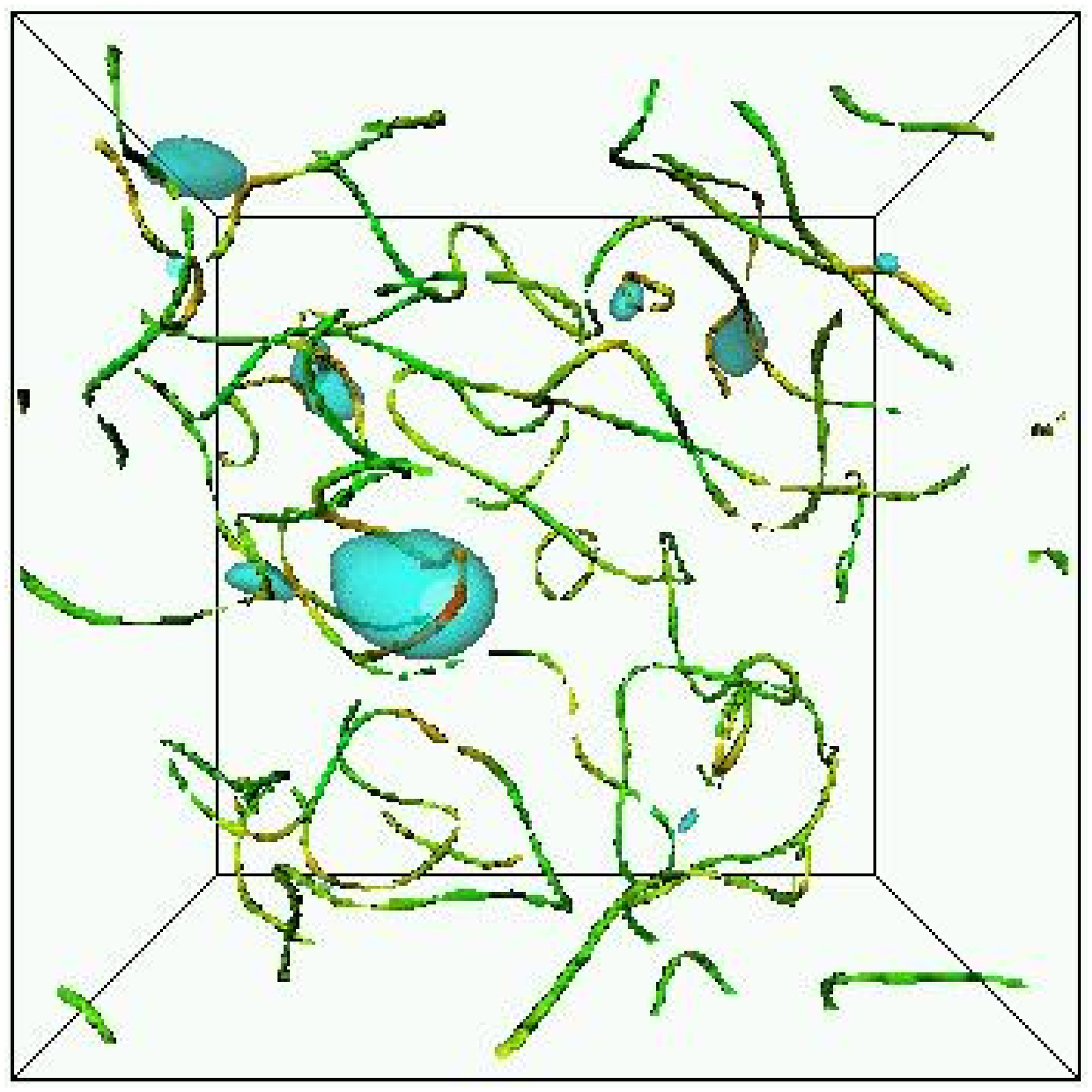}
\includegraphics[width = 0.5\textwidth,clip]{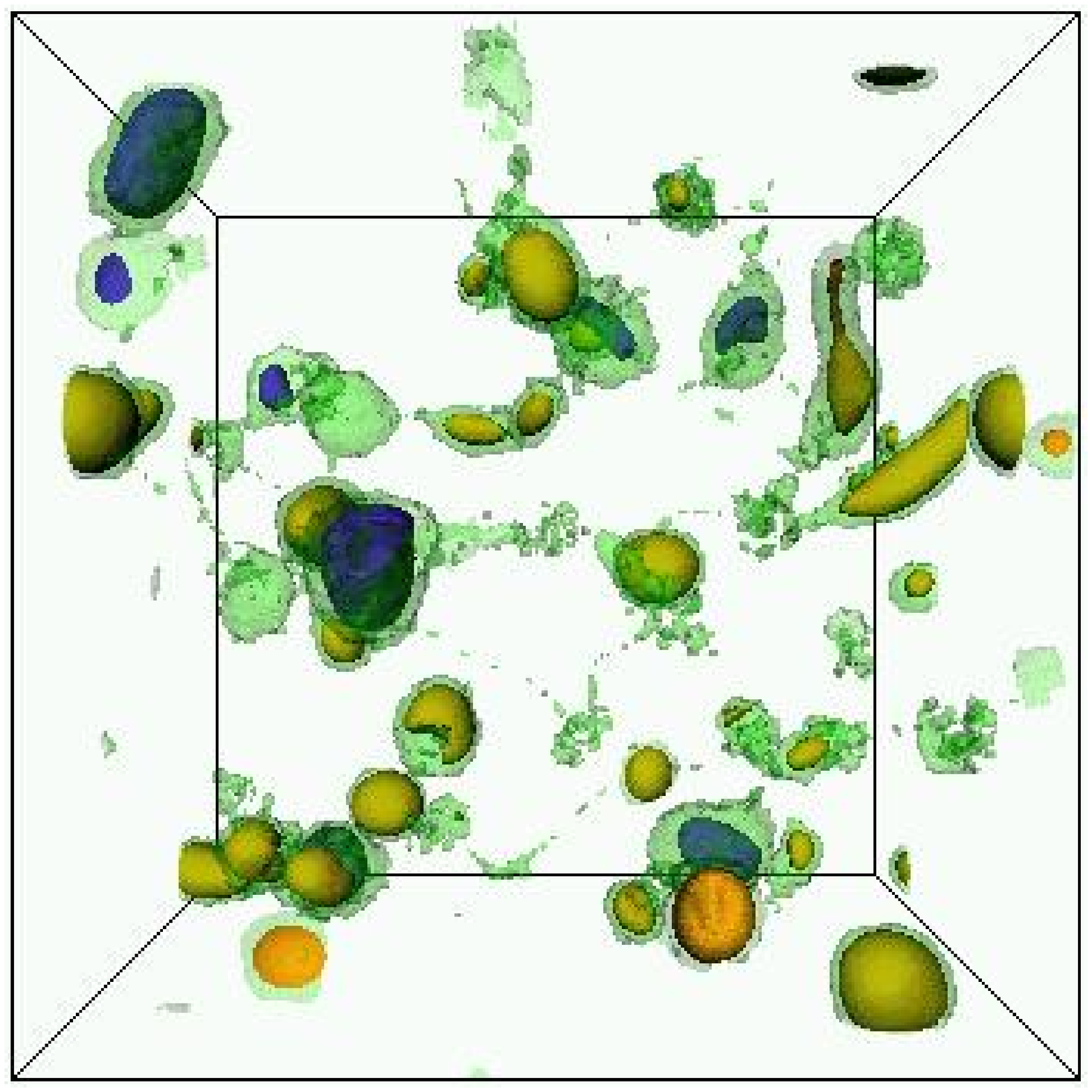}
\caption{Left: transparent surface: $\sg = -\sg_c$ at time $tm = 130$; strings:
$|\vr|^2=v^2/10$ at time for which spatial average $\overline{\sg} = 0$
for the first time. Right: transparent green surface:
$|\vr|^2=v^2/10$; blue: $\sg = \sg_c$; yellow: $\sg=-\sg_c$;
time $tm = 270$. From \cite{Copeland:2002ku}.}
\label{Copeland1}
\end{figure}

Figure \ref{Copeland1} shows surfaces where $\sg$ has fluctuated
to $\pm\sg_c$, and tubes of $|\vr|^2 = v^2/10$, inside of which
the centers of strings are located at $\vr=0$. At the later time
$tm = 270$ the strings have mostly decayed, but not the `hot
spots'. The authors of \cite{Copeland:2002ku} remark that the hot
spots play an important role in the efficiency of tachyonic
preheating.

\subsubsection*{Example: Tachyonic electroweak quench
\cite{Skullerud:2003ki}}
The emergence and development of particle distribution functions
in a tachyonic electroweak transition has been studied with the
SU(2)-Higgs model
\[
-{\cal L}_{\rm SU(2)H} = \frac{1}{2g^2}\,\tr F_{\mu \nu}F^{\mu \nu}
%\right.\\&&\left.\frac{\mbox{}}{\mbox{}}
 +(D_{\mu}\vr)^{\dagger}D^{\mu}\vr
 -\mu^{2}\vr^{\dagger}\vr+\lambda(\vr^{\dagger}\vr)^{2},
\]
with fairly realistic couplings,
%$g^2 = 4/9$,
$m_W=82$ GeV, $\mh/m_W =\sqrt{2}$.
The initial conditions for the Higgs field were taken
to be the quench \eqref{quench}, `just the half', with
$k_{\rm max} = \mu = \mh/\sqrt{2} = 0.25/a_s$. The initial gauge field potentials $\vecA$
were set to zero (since they are on the same footing as
$\vr_\veck$ with $k>k_{\rm max}$), and the canonical conjugates
$\dot\vecA$ then followed from the Gauss constraint in the
temporal gauge ($\vecA_0=0$) used for solving the e.o.m. The
particle distribution functions $n_k$ for the Higgs and gauge
fields were computed as in \eqref{partnumbers}, after fixing the
gauge to `Coulomb' or `unitary'.

\begin{figure}
\be
\includegraphics[width = 0.6\textwidth,clip]{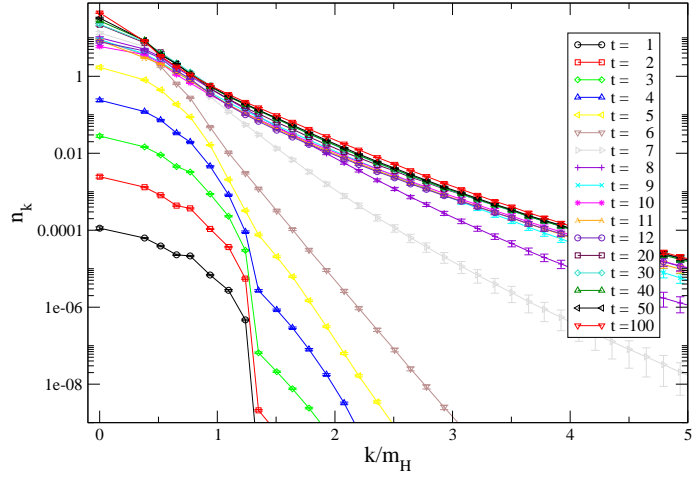}
\ee
\caption{Coulomb-gauge W-particle numbers as a function of momentum, time
in units of $m_H^{-1}$. From \cite{Skullerud:2003ki}.}
\label{Skullerud1}
\end{figure}

Figure \ref{Skullerud1} shows the development of the $W$-particle
numbers. At first the low-momentum modes increase exponentially
fast due to their coupling to the unstable Higgs modes. At time 7
there is a change of behavior in the tail of the distribution.
Approximate thermalization was found to set in from $t \mh = 30$
or so onwards, and not much happened in $50\lesssim t \mh < 100$.
At $t\mh = 100$ there is approximate gauge-independence in the
transverse $W$-particle numbers, but the longitudinal
$W$-modes settle at a slower rate. Effective Higgs and $W$
temperatures at that time agreed reasonably well. Effective
chemical potentials $\mu_{\rm chem} \approx 0.7$ (0.9) for $W$ (H)
are expected to vanish much later.

\subsubsection*{Example: Cold electroweak baryogenesis
\cite{Tranberg:2003gi,Tranberg05}}
In this simulation the scenario was tested using an effective
CP-violating term in the equations of motion following from the
lagrangian
\[
{\cal L} = {\cal L}_{\rm SU(2)H} -
%\kp\, \vr^{\dagger}\vr\, \tr\,F_{\mu\nu}\tilde F^{\mu\nu}
\frac{3\dl}{m_W^2}\, \vrd\vr\,q,
\quad
q= \frac{1}{16\pi^2}\, \tr\, F_{\mu\nu}\tilde F^{\mu\nu}.
\]
Here 3 is the number of families and $\dl$ parametrizes effective
CP violation, perhaps from physics beyond the Standard Model.
Since $q$ is a total
derivative,
$q=\dmu j^{\mu}_{\rm CS}$, it
is customary to rewrite the anomaly equation \eqref{anomaly1} in
the form
\[
B(t) = 3\langle\NCS(t) - \NCS(0)\rangle,
\]
with $\NCS = \intvecx j^0_{\rm CS}$ the Chern--Simons number. A
useful parameter is furthermore $\nw$, the winding number of the
Higgs field, since
$\NCS\approx \nw$, is expected to hold when fluctuations have
sufficiently diminished.
The parameters of the simulation were $m_W= 82$ GeV, $\mh/m_W =1$,
$\sqrt{2}$, $\sqrt{3}$ and 2, quench initial conditions with
$k_{\rm max}= \mu$.
% = \mh/\sqrt{2}$.
The lattice implementation of CP violation leads to {\em implicit}
e.o.m.\ which were solved by iteration at a substantially
increased time of computation.

\begin{figure}
\bc
\includegraphics[width = 0.6\textwidth,clip]{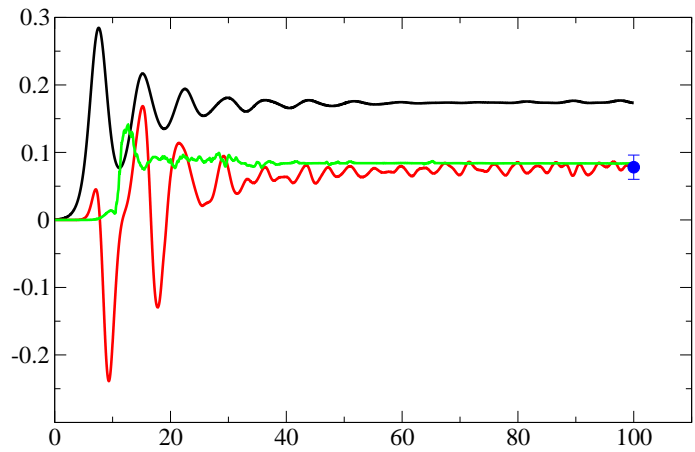}
\ec
\caption{$\langle\vrd\vr\rangle$(black),
$\langle\NCS\rangle$(red),
$\langle \nw\rangle$(green) vs time in units $\mh^{-1}$;
$\dl = 1$,
$m_{\rm H} = \sqrt{2}\, m_W$. From \cite{Tranberg05}.}
\label{Tranberg1}
\end{figure}
\begin{figure}
\bc
\includegraphics[width = 0.7\textwidth,clip]{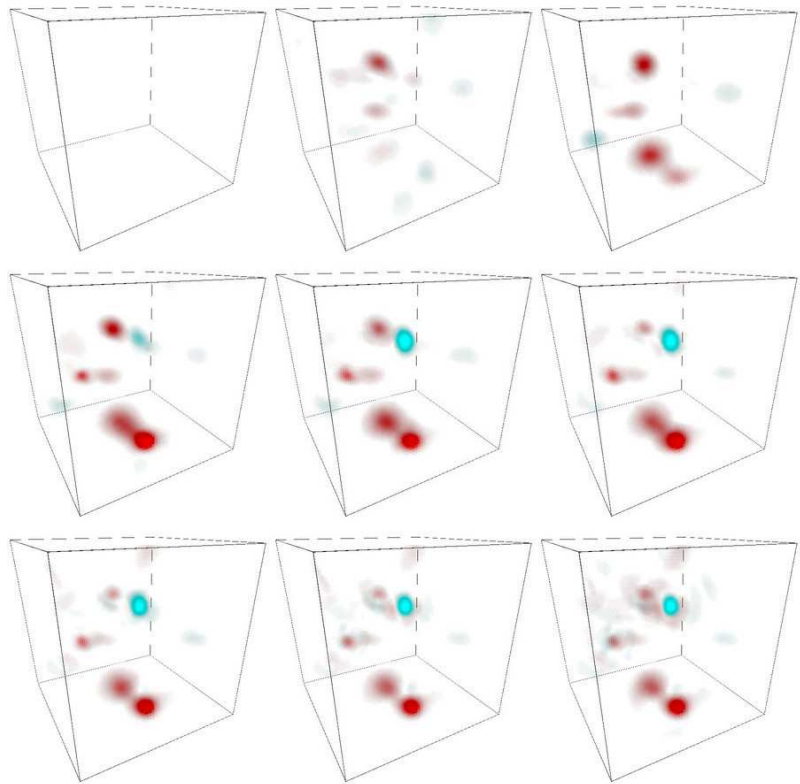}
\ec
\caption{3D plot of Chern-Simons number density for times $t\mh = 7$, 8, \ldots,
15. Red: positive, blue: negative.}
\label{Meulen1}
\end{figure}

Figure \ref{Tranberg1} shows an example of the development of
$\langle\vrd\vr\rangle$, $\langle\NCS\rangle$ and $\langle\nw\rangle$.
We see Higgs length at first rising
exponentially and then executing a damped oscillation around its
finite-temperature equilibrium. The CP violation causes an
asymmetry in the Chern-Simons number, close to the average winding
number, which settles earlier. This leads to a baryon asymmetry
$n_B/n_\gm \approx(4\pm 1) 10^{-5}\, \dl$, which fits observation
for the quite reasonable looking value
$\dl \approx 1.5\;10^{-5}$.

\subsubsection*{Example: winding defects \cite{Meulen}}

A more detailed understanding of the production of Chern-Simons
number might enable us to get more grip on the effects of
different forms of CP violation, e.g.\ by reducing the number of
variables through modelling. This production appears to be seeded
by regions with winding number of $\approx \pm 1/2$, `half-knots'
\cite{Meulen}. Figure \ref{Meulen1} shows an example the development of
the
%winding number density in the Higgs field,
Chern-Simons number density in the tachyonic electroweak quench.
Parameters and initial conditions as in  Fig.\
\ref{Skullerud1}.

\subsubsection*{Example: kinetic turbulence and scaling
 \cite{Micha:2004bv,Diaz-Gil:2005qp}}

Recently it was demonstrated that concepts of turbulence apply
very well to preheating dynamics \cite{Micha:2004bv}. Coefficients
of power laws and scaling relations were derived on the basis of
Boltzmann equations for classical fields, and applied successfully
to results of numerical simulations of an inflaton field
interacting with itself and with another scalar field
\cite{Micha:2004bv}. In a contribution to this meeting, similar
behavior was shown to apply to a tachyonic electroweak transition
\cite{Diaz-Gil:2005qp}. This simulation included the U(1)
hypercharge/electromagnetic field, as it was motivated by
investigating a seeding mechanism for the currently observed
magnetic fields on large scales. Parameters:
$m_W = 6.15$ GeV, $m_{\rm H}/m_W = 4.65$,
$M (t_{\rm i} - t_c) = 5$ (cf.\ \eqref{nonquench}),
$g_{SU(2)}/g_{U(1)}$ as in Standard Model.

\begin{figure}[h]
\includegraphics[width = 0.35\textwidth,angle=-90,clip]{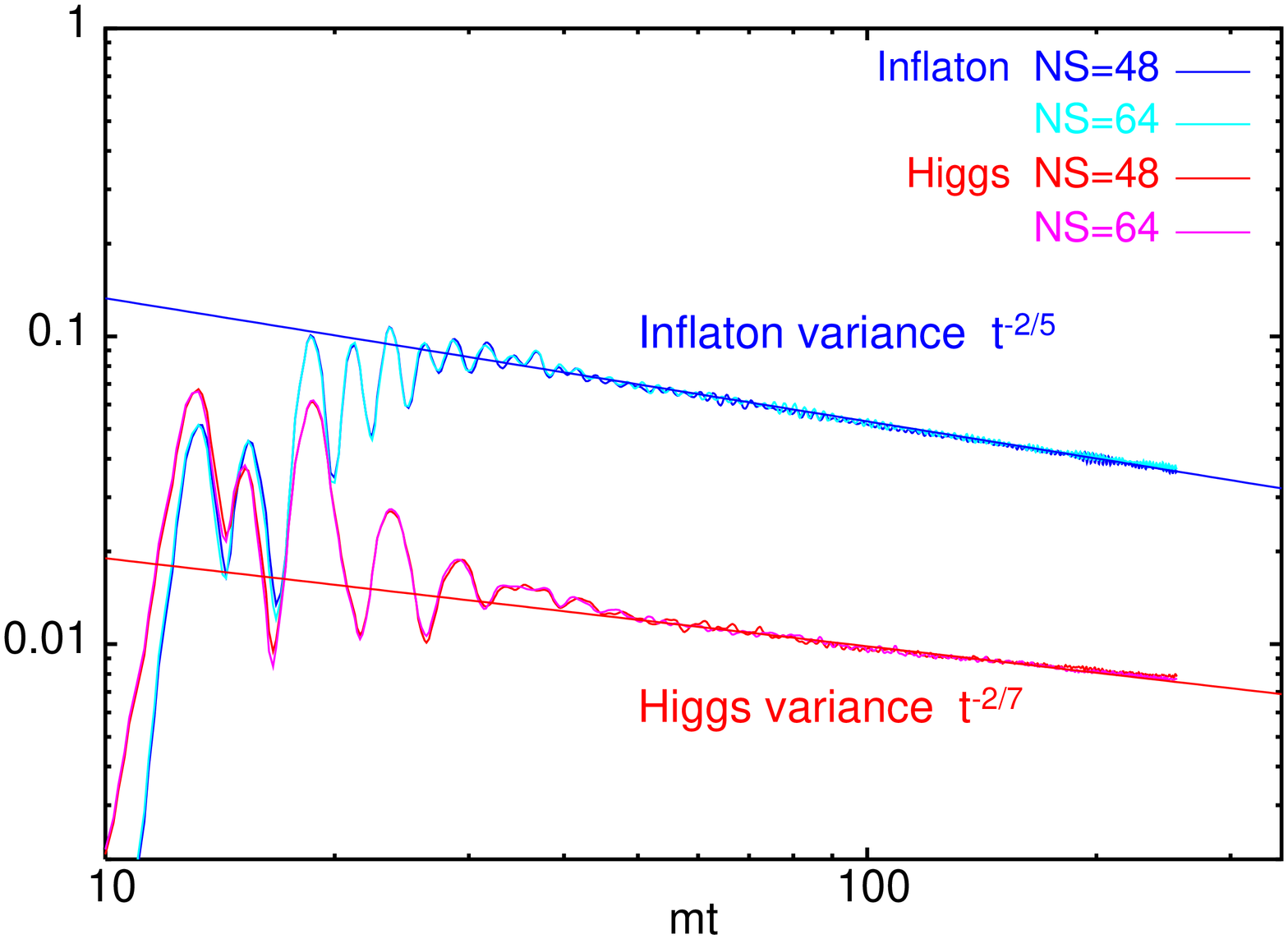}
\includegraphics[width = 0.35\textwidth,angle=-90,clip]{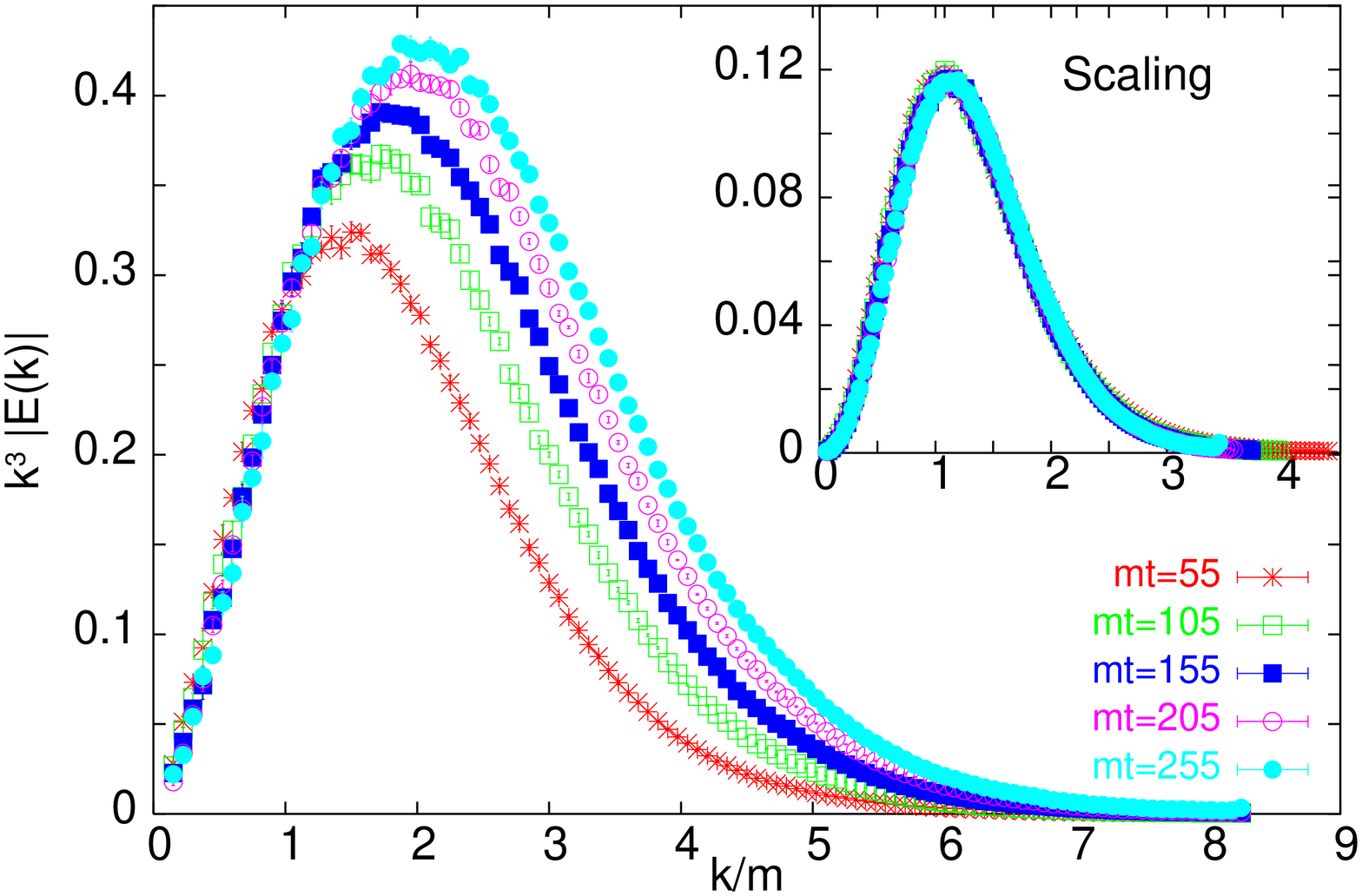}
\caption{Left: turbulent scaling of Higgs $\langle\vr\vrd\rangle$
and inflaton $\langle\sg\sg\rangle$ $\propto t^{-\nu}$, $\nu
= -2/(2m-1)$, $m=3$ ($\sg$), $m=4$ (H).
Right: self-similar SU(2) spectrum: $n(k,t) = t^{-q}\,
n_0(kt^{-p})$,
$q=(7/2) p$, $p=1.1/7$. From \cite{Diaz-Gil:2005qp}.}
\label{GB2}
\end{figure}

Figure \ref{GB2} shows power-law behavior of the variance of the
Higgs and inflaton fields, and scaling of the $W$-particle numbers
(defined here in terms of the energy spectrum).

\section{$\Ph$-derivable approximation}
At larger times the universe is expected to approach local
equilibrium if the dynamics is fast relative to the Hubble rate.
Then the classical approximation breaks down because the energy
becomes re-distributed over all field modes (equipartition),
either leading to lattice artefacts of the Rayleigh-Jeans type, or
causing the effective temperature to vanish in case of no cutoff.
A quantum description is needed to be able to describe true
thermalization through scattering and damping. This has also a
non-perturbative features, e.g.\ a damped oscillation $e^{-\gm
t}\cos(\om t)$ with $\gm = c\lm^2$ misses essential features when
retaining only low orders of the expansion
$e^{-\gm t} = 1 - c\lm^2 + \cdots$.
In diagrammatic language, we need to sum an infinite series of
diagrams, in particular for two-point functions. One possibility
is using the hierarchy of Dyson-Schwinger equations. Recently
so-called $\Ph$-derivable approximations
\cite{Baym:1962sx,Berges:2002wf} have been used in numerical
simulations, and found to be able to capture thermalization
\cite{BeCo01}. It is a real-time effective-action method in which
the two-point function is treated as a basic dynamical variable,
in addition to the mean field. The effective action is written as
\cite{Cornwall:1974vz}
\[
%\Gm[\ph,G] = S[\ph] -\frac{i}{2} \Tr \ln G + \frac{i}{2}
%\Tr \frac{\dl^2 S[\ph]}{\dl\ph\dl\ph}\, G + \Ph[\ph,G]
\Gm(\ph,G) = S(\ph) -\frac{i}{2} \Tr \ln G + \frac{i}{2}
\Tr \frac{\dl^2 S(\ph)}{\dl\ph\dl\ph}\, G + \Ph(\ph,G),
\]
where $S$ is the classical action, $G$ is the two-point function, and
$\Ph$ is a sum of two-particle irreducible (2PI) diagrams,
\be
\includegraphics[width=0.7\textwidth]{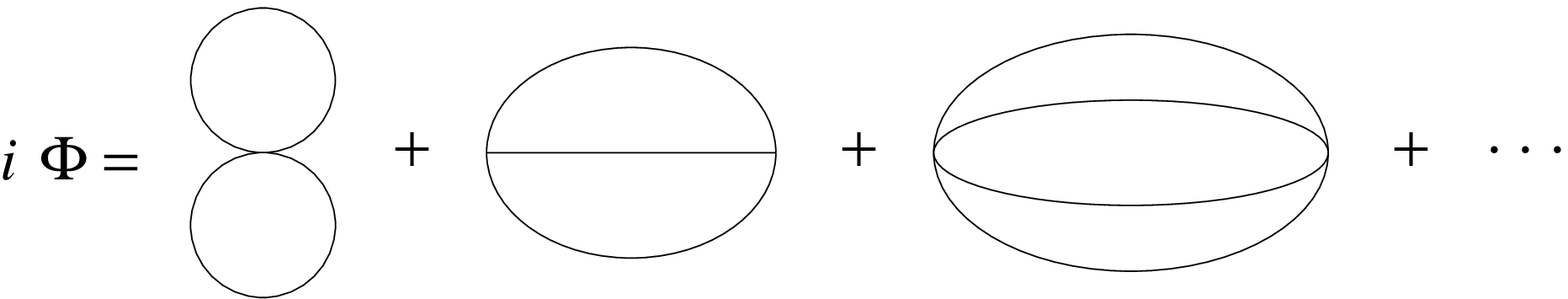}
\label{Phidia}
\ee
with bare vertex functions and dressed two-point functions $G$.
The equations of motion are given by
\[
\frac{\dl \Gm(\ph,G)}{\dl \ph(x)} = 0,
\quad
\frac{\dl \Gm(\ph,G)}{\dl G(x,y)} = 0,
\]
and after separation into contributions with definite
time-ordering, indicated by $>$, they have a non-local form, e.g.\
for $G$:
\bea
\left[ \partial^2 - \mu^2 - 3\lm\ph^2 - 3\lm G(x,x) \right]G^>(x,y) &=&
2 \int_0^{x^0} dz^0\int d^3 z\;
{\rm Im}[ \Sg^>(x,z;\ph,G)] G^>(z,y)
\label{Geom}\\
&&
\mbox{} -2\int_0^{y^0}dz^0\int d^3 z\; \Sg^>(x,z;\ph,G)\, {\rm Im} [G^>(z,y)].
\nonumber
\eea
Here the selfenergy,
%$\Sg(\ph,G)$
\[
\Sg(x,y) = 2i\frac{\dl \Ph}{\dl G(x,y)},
\]
is again given by a series of diagrams,
\be
\includegraphics[width=0.7\textwidth]{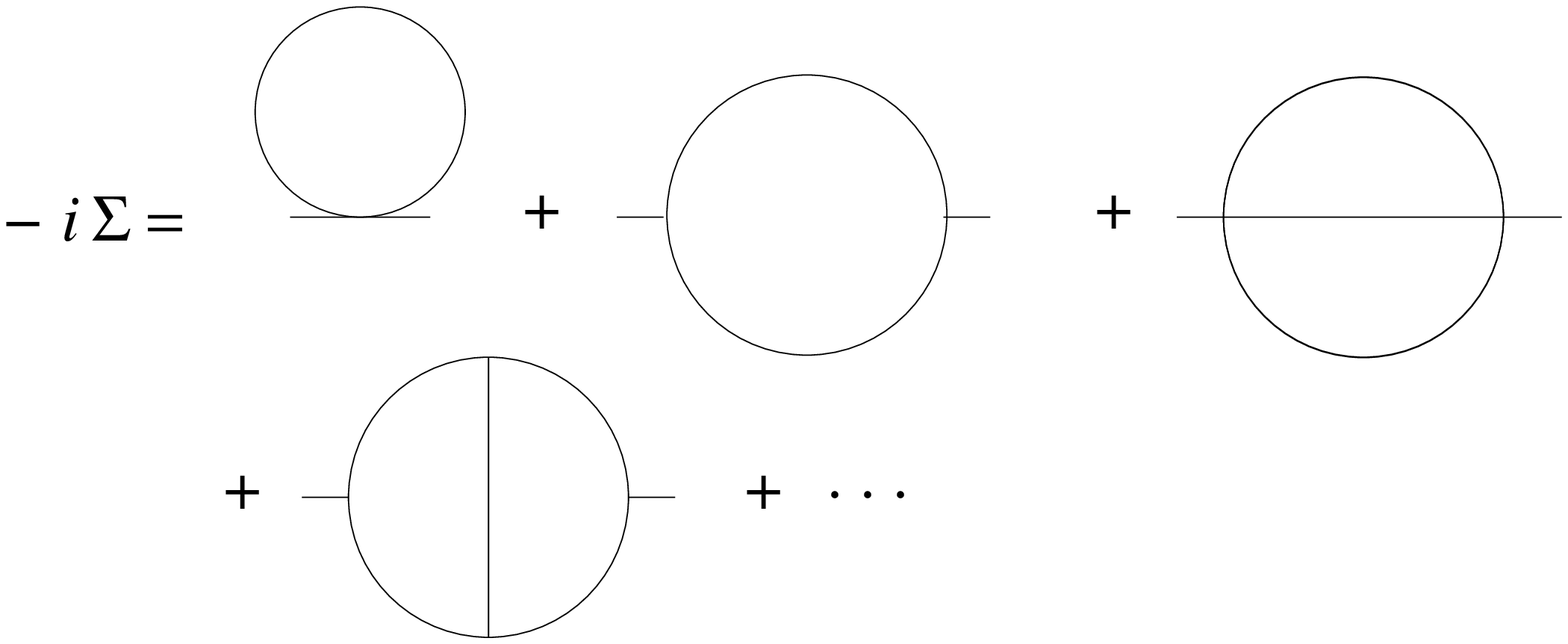}
\label{Sgdia}
\ee
Interesting aspects of $\Phi$-derivable methods are:\\
\indent
- they are `thermodynamically consistent' \cite{Baym:1962sx}\\
\indent
- global symmetries $\to$ Ward-Takahashi Identities, \\
\indent
\mbox{ } e.g.\ translation invariance $\to$ energy-momentum conservation\\
\indent
- they are renormalizable
\cite{vanHees:2001ik,VanHees:2001pf,Blaizot:2003br,Blaizot:2003an,Cooper:2004rs,Cooper:2005vw,Berges:2004hn,Berges:2005hc}

\noindent
$\Ph$-derivable approximations are obtained by
truncating an expansion of $\Ph$, guided by the number of loops,
or the order in a coupling- or $1/N$-expansion
\cite{Berges:2002wf,Aarts:2002dj}. Putting
$\Gm$ on a space-time lattice leads to a numerically viable,
albeit computationally expensive scheme. The non-locality of
\eqref{Geom} in space can be taken care of
(in the usual homogeneous case) by Fourier transformation, which
gives an e.o.m.\ for each Fourier mode. The non-locality in time,
while in accordance with causality, leads to numerically expensive
`memory kernels'. Moreover, these kernels themselves can only be
computed in simple approximations, e.g.\ keeping only diagrams of
the type given in the first line in \eqref{Sgdia}; for example,
the diagram shown in the second line has two internal vertices,
implying two four-dimensional summations over space-time, which is
too expensive, numerically.

Initial conditions are given in terms of $\ph$ and $G$, and the
results of solving the e.o.m.\ can be used directly for computing
expectation values $\langle O  \rangle$ of observables
$O(\ph,G)$ -- there is no need for averaging over initial configurations
as in classical approximations.

\begin{figure}
\includegraphics[width = 0.5\textwidth,clip]{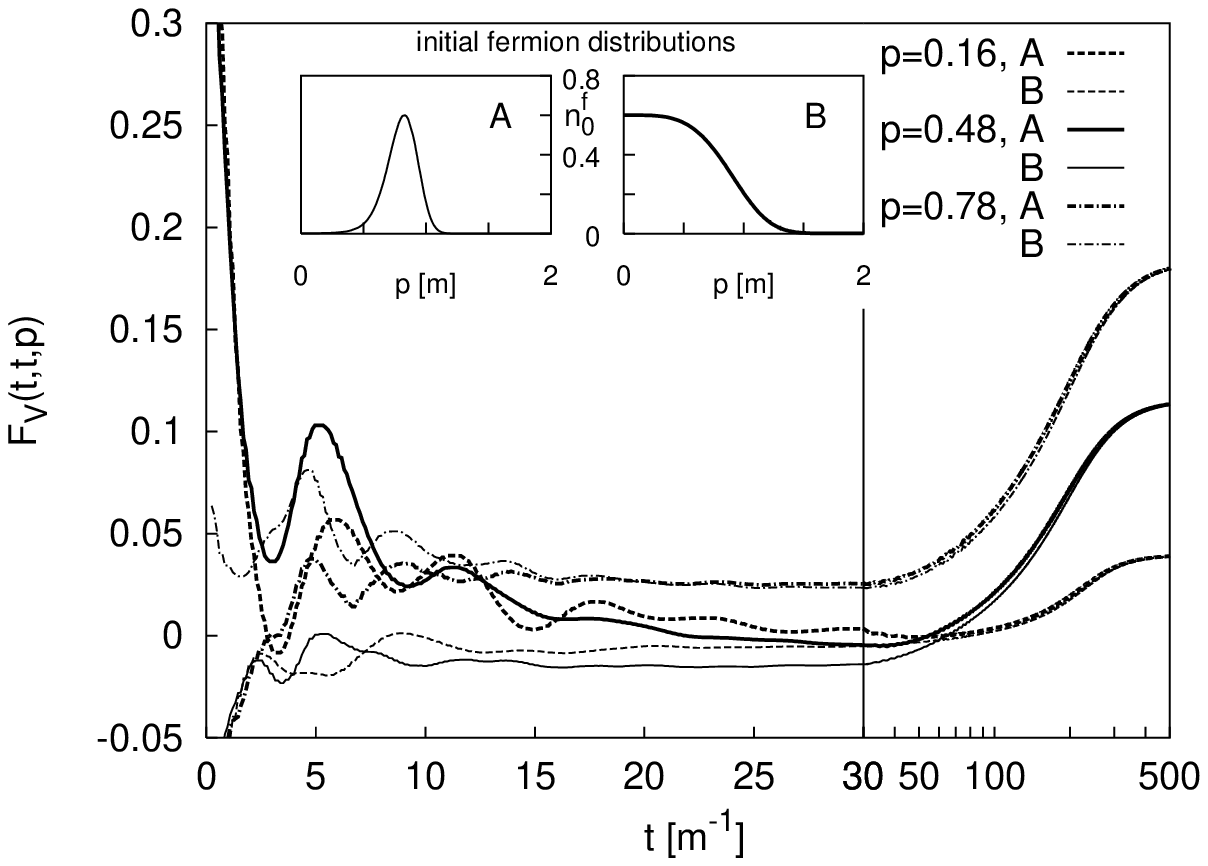}
\includegraphics[width = 0.5\textwidth,clip]{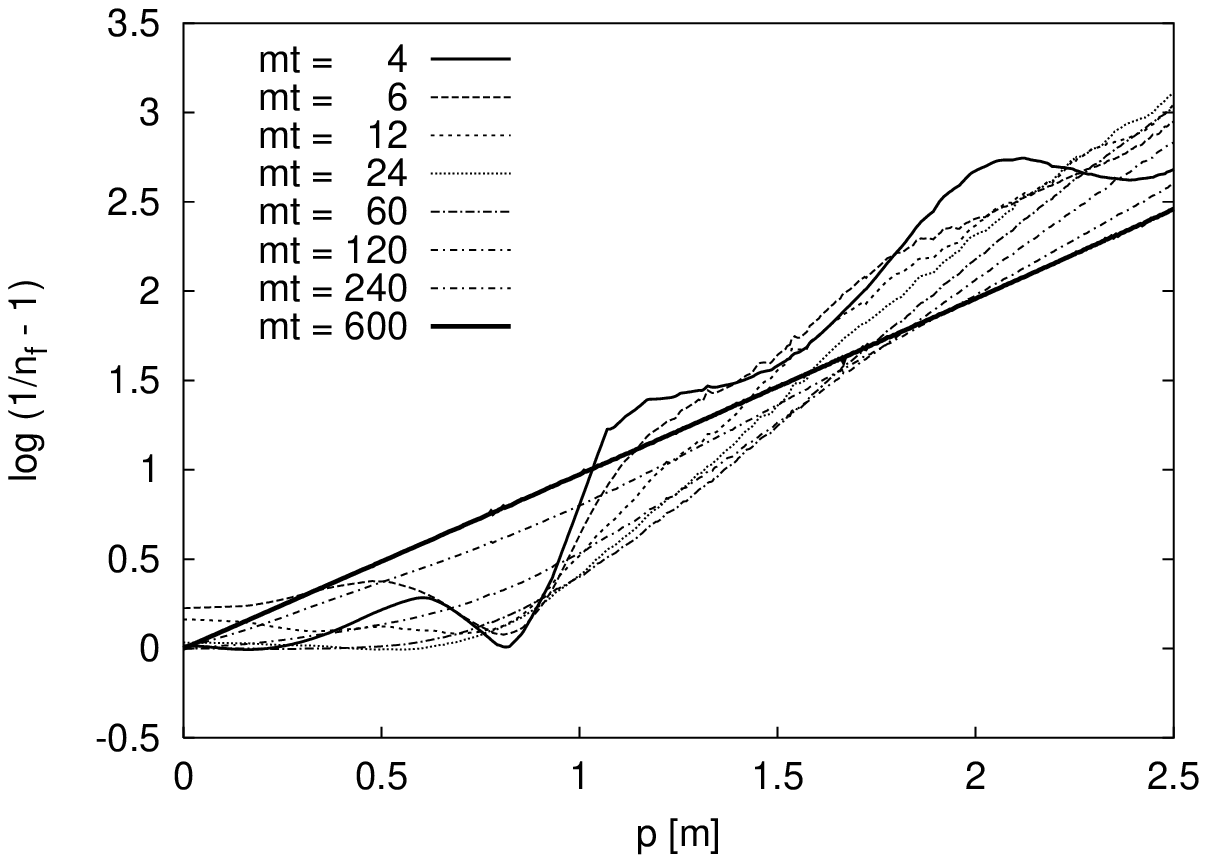}
\caption{Left: Equal-time $F_V(t,\vecp;t,\vecp)$ vs time
for several $p$ and two initial conditions. Right: Approach to the
thermal state:
$\ln(-1+1/n_p)$ vs $p$ at various times. From
\cite{Berges:2002wr}.}
\label{Berges1}
\end{figure}

\subsubsection*{Example: chiral quark-meson model
\cite{Berges:2002wr}}
The lagrangian of the model is of the Yukawa form
\[
- {\cal L}= \psb\gm^\mu\dmu\ps + \thalf(\dmu\sg\partial^\mu\sg +
\dmu\pi^a\partial^\mu\pi^a)
%\\&&\mbox{}
+ \thalf\,\mu^2(\sg^2 + \pi^a\pi^a) + g\psb(\sg +
i\gm_5\ta^a\pi^a)\ps,
\]
where the $\ta^a$ are Pauli matrices, and $\Ph$ is truncated to
the simple diagram
\be
\includegraphics[width=0.2\textwidth,clip]{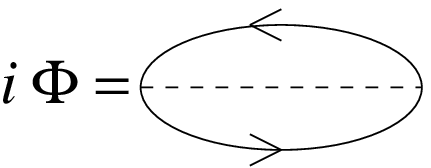}
\label{PhiYuk}
\ee
In ref.\ \cite{Berges:2002wr} the two-point functions are split
into real and imaginary parts,
$G(x,y) \sim F(x,y) -i\rh(x,y)/2$, with spectral functions $\rh$ and
`statistical functions' $F$, and the discretization is done
directly for the e.o.m.\ in terms of components of the fermion
two-point function, as in
\[
\rh = \rh_S+\rh_P\, i\gm_5 + \rh_V^\mu\gm_\mu +
\rh_A^\mu\gm_\mu\gm_5 + \thalf \rh_T^{\mu\nu}\sg_{\mu\nu}.
\]
Remarkably, no fermion-doubling was observed.

Figure \ref{Berges1}(left) shows spatially Fourier transformed
$F_V(t,\vecp;t,\vecp)$ vs time for three values of $p$ and two
different initial conditions. The Yukawa coupling $g=1$. Memory
of the initial condition gets lost, which can be phrased in terms
of a damping time
$1/\gm_{\rm F}^{\rm damp}\approx 15\, m^{-1}$, with $m$
a `thermal mass'. The approach to the final values takes much
longer. Figure \ref{Berges1} (right) shows the fermion
particle-distribution function
$n_p$ at various times in the form
$\ln(-1+1/n_p)$, starting from an initial condition out of
equilibrium. At large times it approaches the line $p/T$,
corresponding to the Fermi-Dirac distribution $n_p =1/
(1+e^{p/T})$ for a massless fermion. So this is clear evidence for
quantum thermalization. Deviations from the FD form due to the
interactions appear to be small, even for a Yukawa coupling as
large as $g=1$. The approach to the FD form can be analyzed in
terms of a thermalization time $1/\gm_{\rm F}^{\rm therm}\approx
95\, m^{-1}$.

%\begin{figure}
%\bc
%%\includegraphics[width = 0.5\textwidth,clip]{join_fermion.eps}
%\includegraphics[width = 0.7\textwidth,clip]{evoldist_FD.eps}
%\ec
%\caption{Approach to the thermal state:
%$\ln(-1+1/n_p)$ vs $p$ at various times.
%From \cite{Berges:2002wr}.}
%\label{Berges2}
%\end{figure}

\subsubsection*{Example: tachyonic `electroweak' quench
\cite{Arrizabalaga:2004iw}}
In this study the scalar sector of the Standard Model is
extended to the $O(N)$ model
\[
%-{\cal L} = \dmu\ph_a\partial^\mu\ph_a/2 -\mu^2 \ph_a\ph_a/2 + \lm (\ph_a\ph_a)^2/4
-{\cal L} = \thalf\, \dmu\ph_\al\partial^\mu\ph_\al -\thalf\,
\mu^2
\ph_\al\ph_\al + \quart\, \lm (\ph_\al\ph_\al)^2,
\quad
\al=1,\ldots,N,
\]
%where $\al=1,\ldots,N$ and
where $\lm\propto 1/N$. The functional $\Ph$ is approximated to
Next-to-Leading-Order in the $1/N$ expansion
\cite{Berges:2002wf,Aarts:2002dj},
after which $N$ is set equal to the SM value 4.

\begin{figure}
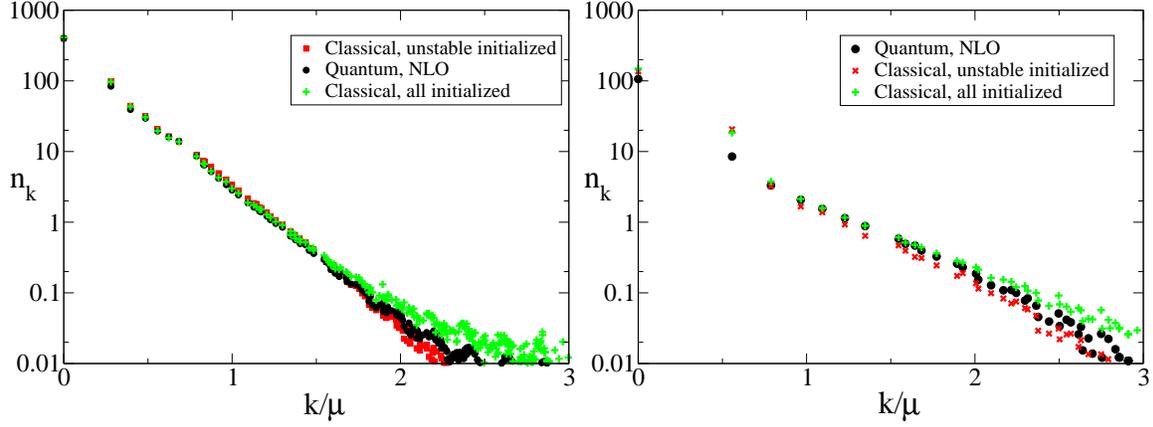

\includegraphics[width = 0.5\textwidth,clip]{Choice1.eps}
\includegraphics[width=0.5\textwidth,clip]{Qvcl.eps}
\caption{Left: particle numbers at time $t\mu=14$;
%NLO(black), only unstable(red), all(green);
$\lm =1/24$ ($m_{\rm H} = 71$ GeV),
$L\mu = 22.4$, $a_s\mu = 0.7$. Right:
similar for stronger coupling
$\lm = 1/4$ ($m_{\rm H} = 174$ GeV), but smaller
volume $L\mu = 11.2$. From \cite{Arrizabalaga:2004iw}.}
\label{Arriza1}
\end{figure}

Figure \ref{Arriza1} shows a comparison with the classical
approximation for the same model with $N=4$ (without $1/N$
expansion). The initial conditions correspond to the quench
\eqref{quench}, so vacuum initial two-point function $G$ in the
quantum $\Ph$-derivable approximation, and `just the half' in the
classical approximation, with two choices of
$k_{\rm max}$ in (\ref{gaussensemble}): $k_{\rm max} = \mu$
(`unstable initialized') and `cutoff' (`all initialized'). It is
reassuring that at time $t\mu = 14$ the two classical
approximations still agree nicely with the $\Phi$-derivable one.
(Such confirmation has been found earlier in 1+1 dimensions
\cite{Aarts:2001yn}.) This agreement has diminished at time
$t\mu
=100$, and it reduces further as time progresses, although the
general shape of the classical spectra remains similar to the
quantal for many hundreds of $t\mu$. Taking the
$\Ph$-derivable approximation as a benchmark, the $k_{\rm max} =$
`cutoff' implementation of the classical approximation
appears to be somewhat more accurate in these plots.
%Note that the linear size in Fig.\ \ref{Arriza2} is half
%that of fig.\ \ref{Arriza1}.

%\begin{figure}
%\includegraphics[width=0.5\textwidth,clip]{Qvcl.eps}
%\includegraphics[width=0.5\textwidth,clip]{Choice2.eps}
%\caption{Particle numbers at time $t\mu=100$ (Left) and 700 (Right);
%%NLO(black), unstable(red), all(green);
%$\lm = 1/4$ ($m_{\rm H} = 174$ GeV), $L\mu = 11.2$.
%From \cite{Arrizabalaga}.}
%\label{Arriza2}
%\end{figure}

\section{Conclusion and outlook}
Before putting a quantum field out of equilibrium on the computer,
approximations had to be made.
%\footnote{However, see \cite{Berges:2005yt}}.
%
Classical approximations can be well motivated in a limited time
regime and they have produced many interesting results. The
dominating classical modes have a minimum wavelength, which makes
it feasible to include topological observables such as winding and
Chern-Simons number densities, and even corresponding terms in the
equations of motion, that are notoriously difficult to deal with
in the quantum theory. We had no time to go into the question of
the size of topological defects in gauge theories
\cite{Hindmarsh:2000kd}. More work is needed on non-abelian gauge fields
without Higgs fields, in view of the issues raised in
\cite{Moore:2001zf}.

In the quantum domain
$\Ph$-derivable approximations appear to be working well also for late
times, and the expected features of quantum thermalization have been observed
in scalar and Yukawa models.
The implementation of Nambu-Goldstone bosons is still an issue
\cite{vanHees:2002bv,Ivanov:2005bv,Berges:2005hc,Cooper:2005vw}.
Gauge theories have to my knowledge
not been simulated yet with this method. In case of three-point interactions
(generic in gauge theories and also in scalar field
models with non-zero mean field),
the two-loop selfenergies have contributions that are prohibitively
numerically expensive. There is the problem of
dependence on the gauge-fixing parameter. It has been argued that
such gauge dependence is acceptable as part of the approximation
\cite{Arrizabalaga:2002hn}.
For non-abelian gauge theory without Higgs fields,
the growth of the coupling in the infrared makes it hard to find good
truncations.

With fermions there are the usual issues associated with doubling
in lattice simulations \cite{AaSm99,Aarts:1999zn}, although this
may be avoidable to some extent in non-gauge theories
\cite{Berges:2002wr}. Perhaps the need to fix the gauge is a blessing
in disguise for chiral gauge theories. Real-time simulations with
fermions might provide an alternative treatment of finite-density.

Finally, in some cases the Hubble expansion effectively causes the
lattice spacing to grow in time, which puts limits on the
meaningful duration of the simulation. And here is still lots to
do in numerical simulations of the Big Bang.

\section*{Acknowledgements}
Many thanks to my collaborators in producing some of the results
presented here, Alejandro Arrizabalaga, Jon-Ivar Skullerud, Anders
Tranberg and Meindert van der Meulen. This work was supported by
FOM/NWO.

\bibliography{lit}

%\begin{thebibliography}{99}
%  \bibitem{...} ....
%\end{thebibliography}

\end{document}